\documentstyle[twocolumn,eqsecnum,aps]{revtex}
\def\gtorder{\mathrel{\raise.3ex\hbox{$>$}\mkern-14mu
             \lower0.6ex\hbox{$\sim$}}}
\def\ltorder{\mathrel{\raise.3ex\hbox{$<$}\mkern-14mu
             \lower0.6ex\hbox{$\sim$}}}

\begin{document}
\draft
\title{Testing gravity in Large Extra Dimensions using Bose-Einstein Condensates
\footnote{This essay received an "honorable mention" in the Annual Essay
Competition of the Gravity Research Foundation for the year 2002.}
}
\author{Steinn Sigurdsson}
\address{525 Davey Laboratory,
Department of Astronomy \& Astrophysics,\\ 
Pennsylvania State University, University Park, Pa 16802\\
steinn@astro.psu.edu
}
\maketitle
\begin{abstract}
Recent conjectures that there are mesoscopically ``large'' extra dimensions,
through which gravity propagates have interesting implications for
much of physics.
The scenario implies gross departures from Newton's law of gravity
at small length scales. 
Testing departures from Coulomb's law on sub-millimetre scales
is hard.

It is now possible to routinely create Bose-Einstein condensates with
de Broglie wavelengths of order a $\mu m$ and total size of order $10 \mu m$.
BEC condensates move coherently under gravitational acceleration, and I propose
that the transverse fringe shift due to the acceleration of pair of interfering BECs 
passing a dense linear mass may be measurable, and provide direct evidence
for anomalous gravitational acceleration. Ideally such experiments are best
carried out in free fall to maximise the time spent by a BEC in the non-Newtonian
regime.

\end{abstract}



\narrowtext
\section{
}
\label{sec:level1}

Recent conjectures have postulated that the hierarchy problem
in physics may be resolved if two (or more) of the extra dimensions
postulated by extensions of the standard model of particle theory,
are compactified on mesoscopic scales - with effective radii
much larger than the Planck scale \cite{ark98,ras99}.
A particularly interesting possibility is in the case of $n=2$
mesoscopic compactification, in which case the implied scale, $R_c$,
for the extra dimensions is of the order of 0.1 mm.
In the simplest theory, the standard model gauge fields are
restricted to (or near) the 3--dimensional brane on which we
live, and only gravity propagates into the bulk of the large extra dimensions (LEDs).
The resulting theory has a number of interesting physical implications 
including possible resolution of the ``hierarchy problem'' and anomalous cross-sections
for particle productions at moderately high energies ($\gtorder 10 TeV$) \cite{Ant98,ark99,kub01}.
There may also be astrophysical implications \cite{sig01,cas00}.

An immediate implication of LEDs is that Newton's law fails on small scales, and
is replaced by an effective potential gradient

\begin{equation}
\nabla \Phi(r) = -(n+1){ {m_1m_2}\over {M_{pl-n}^{n+2}} } { {1}\over {r^{n+2}} } \quad r \ll R_c
\end{equation}
where $n$ is the number of large extra dimensions, and $M_{pl-n}$ is the higher
dimensional Planck mass, implying an effective 
4-D Planck mass $M_{pl} \sim M_{pl-n}^{1+n/2} R_c^{n/2}$ \cite{ark98}.
The laboratory experimental
constraints on deviations from Newton's law on scales less than 1 cm \cite{lcp99} are
very weak, so the
conjecture is not directly excluded by direct experiments, although experiments in
progress will either detect the predicted deviation, or constrain $R_c$
(or equivalently $M_{pl-n}$). If correct, LEDs have many implications for physics
on different scales, some of which will be tested in the near future. 
Experiments to directly measure deviations from Newtonian gravity on sub-millimetre scales are underway
(see \cite{web00}).

It is now possible to routinely generate, in the laboratory, Bose-Einstein condensates
with de Broglie wavelengths of order a micron and total sizes of order $10 \mu m$, and
to manipulate and transport coherent ensembles of $\gtorder 10^6$ 
ultra-cold atoms \cite{becrev1,becrev2,becrev3,atomchip}. Atom interferometry can be used
to obtain both high precision measurements of absolute gravitational acceleration and
gravitational gradients \cite{pet01,sad98}. Also, clearly, the dynamics of a Bose-Einstein
condensate are affected by gravitational forces \cite{gerb01}.

A linear cylindrical mass, with length $l (\gg R_c)$, produces a transverse gravitational acceleration
$g_T = 2G\mu /r$ for $r > R_c$, where $\mu = \pi \rho a^2$ is the mass per unit length
of a cylinder density $\rho $ and radius $a$. As a simple example, consider a thin walled
hollow glass cylinder, with a segment length $l$ filled with a high density substance
(such as mercury, lead or gold), and another comparable length segment empty. If we conjecture that 
$R_c < 0.1 $mm, but $R_c > 0.01$ mm, then for $a = 10 \mu m$ the gravitational field 
close to the cylinder is non-Newtonian with $g_T \approx 2 G \mu /r^3 $.
So the transverse acceleration close to the filled cylinder may be two orders
of magnitude larger than in the Newtonian case. Using a cylinder with a filled core
and a contiguous empty core segment would allow ``blind tests'' of anomalous
transverse acceleration by sliding the cylinder so that either the empty
or the dense core were in the path of the BEC pair. The homogenous structure would hopefully 
lead to consistent systematic offsets
due to surface effects and other non-gravitational perturbations.

A Bose-Einstein condensate falling past such a cylinder will experience a transverse
impulse. Falling under gravity, the BEC will have vertical speeds of order cm/sec,
and traverse the conjectured non-Newtonian regime in a few milliseconds. The resultant
transverse impulse $\Delta v_T = \int - \nabla \Phi dt \sim 4 \times 10^{-10} $ cm/sec.
The resultant Doppler shift would be of the order of $\mu$Hz for reasonable atomic
transitions, which is not practical for direct detection.
If a pair of BEC condensates is sent past the cylinder on either side the resultant
transverse shift will be a order 10 femtometres. Launching the BEC pair vertically
upwards will roughly double the shift and eliminate some sources of systematic
error. With a de Broglie wavelength of a micrometer, measuring the phase shift
due to the presence of cylinder with dense core, as opposed to an empty cylinder
would require measuring the phase accuracy to $O(1/n)$, where $n$ is the number
of atoms in the BEC, which can be done in principle \cite{orz01}.

Additional precision may be obtained by doing multiple vertical traverses of the system.
However, the limiting factor above is the high vertical speed of the BEC as it falls under
gravity. If the speed of the BEC crossing the putative non-Newtonian regime is smaller, 
the transverse impulse is proportionately larger and the transverse phase shift is also
proportionately large as the BEC can undergo transverse coasting due to the
anomalous transverse impulse for longer (up to the
intrinsic lifetime set by the ballistic expansion of the BEC). 
Conducting such an experiment in orbit, for example on the International Space Station,
might produce phase shifts four orders of magnitude larger, and correspondingly easier to measure.
The resultant transverse
velocity would also be correspondingly larger, and might conceivably be detectable,
particularly if the BEC is allowed to traverse past a an array of multiple dense cylinders.

Interferometry of coherent mesoscopic atomic ensembles may be used to
detect anomalous acceleration due to non-Newtonian gravity on sub-millimetre scales.
A simple experimental setup of a thin cylinder with dense core, and a contiguous control segment
with empty core should produce anomalous transverse velocity shifts in pairs of
Bose-Einstein condensates leading to fringe shifts upon the subsequent interference by the pairs, 
that are measurable in principle.
The experiment is best done in free fall, since the gravitational acceleration on
Earth's surface limits the coasting time in the setup postulated here.


\end{document}